# Optically Induced Aggregation In Single Walled Carbon Nanotubes Functionalized with Bacteriorhodopsin

Astha Sharma[1], E. Senthil Prasad[2] and H. Chaturvedi[3*]

[1] *Dept. of Physics, Fergusson College, Pune, India.*

[2] *Institute of Microbiology Technology (IMTECH), Chandigarh, India*

[3] *Dept. of Physics, Indian Institute of Science Education & Research (IISER), Pune, India*

We report optically induced aggregation and subsequent separation of selective single-walled carbon nanotubes (SWNT) functionalized with bacteriorhodopsin in the aqueous solution. Well-dispersed, aqueous solutions of hydrophobic pristine SWNT were prepared using a biocompatible surfactant. Dispersed SWNTs were then functionalized with biologically synthesized, optically active purple membrane from Halobacterium salinarium S9. Bacteriorhodopsin is the optically active protein of the purple membrane. Charge transfer and interactions between an optically active purple membrane (PM) and nanotubes affect the stability of dispersion. Enhanced aggregation in these well-dispersed, stable solutions of SWNT were observed under a lamp with broadband visible frequency. Concentration of SWNTs shows rapid, optically induced aggregation of over 70% in 4 hours. This enhanced rate of aggregation under light was further investigated using specific band pass filters. The rate of aggregation was found to depend on the absorption band of the optically active PM. Raman spectra of the optically separated, bio-nano hybrid complexes show stable, preferential binding between the optically active PM and SWNTs of specific diameters.

Bio-nano hybrid complex are materials of active interest due to its immense potential in diverse applications such as bio-mimic devices, sensors, electro-optics, biotechnology, cancer therapy and diagnostics etc.[1-4] Biomolecules, specifically an optically active purple membrane (PM) have long been studied for photonic and electro-optic applications. Bacteriorhodopsin (bR) is the optically active center of the PM which functions as a photo-induced proton pump by converting optical energy into proton gradient across the membrane.[5] bR amounts to 75% of the PM by mass, while the other 25% is due to lipids. Bacteriorhodopsin is a photosensitive seven helix transmembrane protein usually found in the halophilic bacteria, Halobacterium salinarium. It is a rare molecule found naturally in crystalline form with size ranging from 400 nm – 800 nm and thickness of 5-10 nm.[6] Crystalline structure provides required chemical and thermal stability to the inner bacteriorhodopsin surrounded by lipids. Photocycle of the bR has been under investigation to study proton pumping mechanism. Due to the intrinsic properties of the PM which include high thermal stability, ability to form thin films with excellent optical and proton transport properties, the PM is proposed as an excellent optical material with myriad of applications.[7]

---

* *Corresponding author: hchaturv@iiserpune.ac.in*

Due to unique optoelectronic properties, one dimensional single walled carbon nanotubes (SWNT) are also considered to be promising material for various electro-optical, photonic devices and biosensors.[8-10] Stable donor-acceptor systems, based on optically active bio-nano hybrid complex of PM and SWNT, may play an important role in the development of next generation cutting edge technologies. Extended delocalized systems of conjugated $Sp^2$ carbon atoms organized in hexagonal arrays, high surface areas, and high aspect ratios provide for ballistic transport in one dimensional SWNT. Stable, functional donor-acceptor systems based on functionalized SWNT are being envisioned for various applications. Moreover, solutions of SWNTs provide opportunity to study interesting effects in colloidal solution of one dimensional particles. Stability and controlled assembly are critical challenges for colloidal systems and "bottom-up" assembly of nanoparticles in solution. This exertion describes optically induced aggregation of pristine SWNT of selective diameters, preferentially functionalized with purple membrane. A fundamental understanding of these processes should lead to a better stability, directed assembly and controlled charge transfer in donor-acceptor systems based on SWNTs.

SWNT functionalized with optically active molecules, electron donating polymers and dyes have been well reported; [11-13] however, the low solubility of SWNTs in aqueous dispersions limits its applicability, especially in functionalization using bio-molecules. This limitation is overcome using bio compatible amine based surfactants. Stable aqueous dispersion of SWNTs has been formed using well established protocol of non-covalent functionalization by PL-PEG.[14] Ultrahigh Purified SWNT were purchased from NanoIntegris and were used as such without any further purification. Stable dispersions of the SWNTs were made in water using 1,2-Distearoyl-sn-glycero-3-phosphoethanolamine ($C_{41}H_{82}NO_8P$) powder, PL-PEG amine as a surfactant, which was used as purchased from Avanti, Polar Lipids, Inc. The stable SWNT dispersion was prepared by the sonication of pristine, hydrophobic nanotubes in aqueous solutions of above mentioned amphiphilic polymer. The hydrophobic lipid chains of PL-PEG amine are strongly anchored on the nanotube surface, whereas the hydrophilic amine chain provides SWNT with water solubility and biocompatibility. Prepared aqueous solutions of SWNT provides for new functionalities to be added to the nanotubes, using bio-molecules such as bacteriorhodopsin. An aqueous solution of SWNTs showed excellent stability for weeks.

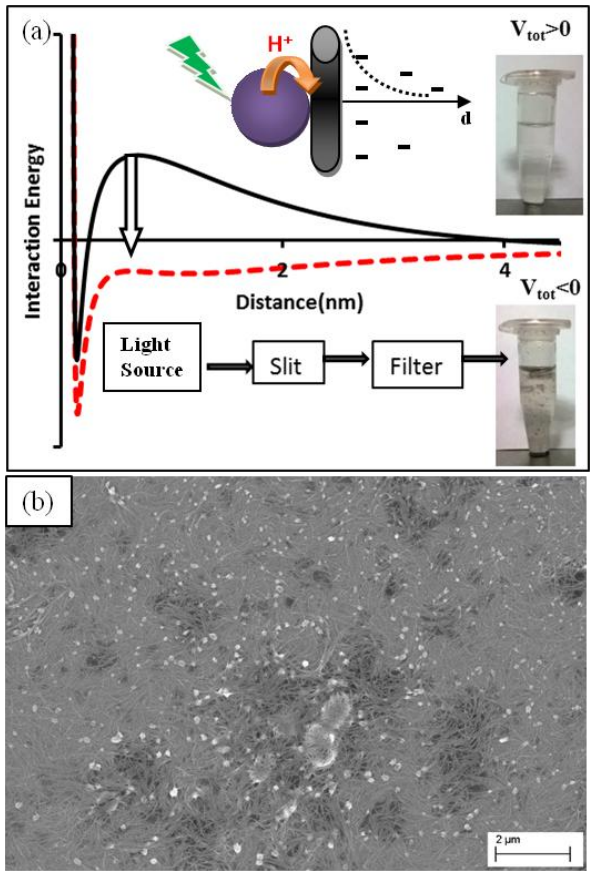

FIG 1 (a) Conceptual model and experimental set up of optically aggregated PM functionalized SWNTs. Interaction and inhomogeneous surface charges leads to aggregation due to decrease in repulsive potential. (b) SEM of aggregated PM functionalized SWNTs showing random surface binding of PM with SWNTs.

Optoelectronic and thin film applications demand high quantity (several milligrams) of the PM with better optical properties.[15] Halobacterium cultivation was performed in a 7 liter photo-bioreactor to give high yield of 14.4 mg/l. 79.36 micro molar (uM) from the stock protein was used for further functionalization with SWNTs. SWNTs tend to coagulate, when they interact steadily with the PM over 24 hours, in dark. Prepared aqueous solution of PM-SWNT was irradiated horizontally through a slit, by broadband mercury white lamp (New port, oriel instruments, USA) for time periods in ratio of [1:2:3:4] starting with 1 hour. Corresponding to each of the samples kept under light exposure, samples of the same concentrations were kept in the dark for a respective duration of time. Fig. 1 (a) illustrates the experimental set up and conceptual model for our observation of optically induced aggregation of the functionalized SWNT dispersions. Pristine dispersions of SWNT do not show any change in concentration over the course of these experiments. Although, PM functionalized SWNTs coagulate very slowly in the dark; optically induced rapid aggregation is observed for the PM functionalized SWNTs, kept under broadband illumination.

DLVO theory is traditionally used to understand stability in colloidal solutions. Even with its inherent limitation due to assumptions on size, charge dielectric properties and dimensions of interacting particles, it is widely being used to understand fundamental interactions in solutions of nanoparticles.[16, 17] Bhattacharya et al[18] recently asked if light has the ability to aggregate particles. This is an interesting question of wider interest. Although optical effects in solutions of gold nanoparticles and functionalized SWNT has limitedly been reported, the phenomenon is not well understood.[19, 20] Here, we attempt to provide theoretical frame work for the proposed conceptual model of our experimental observations. According to the theory, total interaction potential ($V_{tot}$) of two spheres with radius $R_1$, $R_2 >> D$, the inter-particle distance is given by $V_{tot} = \left(\frac{-A_H}{6D} + Ze^{-kD}\right)\left[\frac{R_1 R_2}{R_1+R_2}\right]$. Here, the first term is due to the attractive Van der Waals force which depends on the Hamaker constant $A_H$. Hamaker constant is related to the dielectric properties of the system. For $\varepsilon_1$, $\varepsilon_2$, $\varepsilon_3$ being the dielectric constants and $n_1$, $n_2$, $n_3$ are refractive indexes of SWNTs and medium (water) respectively, $A_H \approx \frac{3}{4} K_B T \left[\frac{\varepsilon_1-\varepsilon_3}{\varepsilon_1+\varepsilon_3}\right]\left[\frac{\varepsilon_2-\varepsilon_3}{\varepsilon_2+\varepsilon_3}\right] + \frac{3hV_e}{8\sqrt{2}} \times \frac{(n_1^2-n_3^2)(n_2^2-n_3^2)}{\sqrt{(n_1^2+n_3^2)}\sqrt{(n_2^2+n_3^2)}\left[\sqrt{n_1^2+n_3^2}+\sqrt{n_2^2+n_3^2}\right]}$. Dielectric response of optically excited PM functionalized

SWNT is expected to be different than of a pristine SWNT, specifically at resonant, absorbing frequencies. A change in the dielectric response should lead to consequent changes in $A_H$. Moreover, the second term in the total interaction energy is repulsive potential due to electrostatic double layer depending on the interaction constant Z.[16, 21] This repulsive Interaction constant 'Z' significantly depends on surface charges, as calculated using equation $Z = 64\pi\varepsilon_r\varepsilon_0 \left(\frac{K_BT}{e}\right)^2 \tanh^2\left(\frac{ze\varphi_0}{4K_BT}\right)$, where $\varepsilon_r$ is the dielectric constant of the solvent. One of the significant assumptions of DLVO theory is a uniform distribution of charge, which is very well applied for systems such as synthetic latex particles. However, interactions between non-uniformly charged surfaces are complex and correspondingly complicated. Theoretical studies show colloidal instability and aggregation of particles as a consequence of charge non uniformity on colloidal interactions.[6] SEM images in Fig. 1 (b) and AFM/TEM images of SWNTs functionalized with PM,[22] show random side-wall binding on SWNTs and conformal changes in adsorbed PM. Non uniformity in surface charge due to random surface functionalization by an asymmetrically charged PM bound to a one-dimensional nanoparticle, is expected to play a critical role in stability of the solution. PM is formed of optically active bR, along with asymmetrically located charged lipids. Upon optical excitation (500nm-600nm), photoisomerizaton causes conformational changes in the bR protein, by initiating proton pumping along the membrane and against the electrochemical gradient.[23, 24] Photoisomerizaton occurs in few picoseconds upon photo cycle initiation which also brings about charge redistribution inducing a photo voltage with nanoseconds of illumination.[24] Light scattering experiments also indicate curving and conformal changes in an excited PM. Conformal changes and random surface binding of PM on SWNTs, predictably results in a complex system with asymmetric charge distribution. Optical excitation in such a system should affect the repulsive double layer because of enhanced charge non uniformity of surface charges at local binding sites. Hence, we believe colloidal instability and aggregation is caused due to cumulative effect of changes in attractive $A_H$ of PM functionalized SWNT and decrease in repulsive potential because of local charge regulation and interaction between SWNT and optically excited PM.

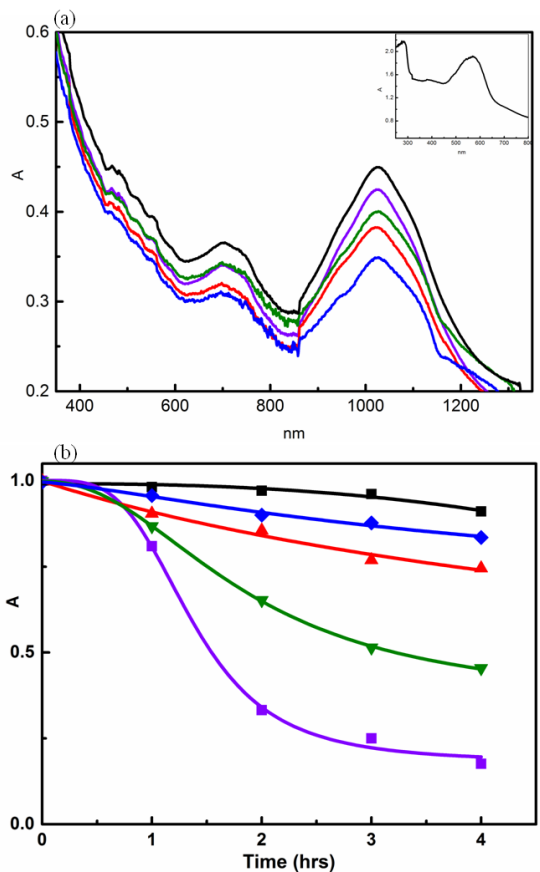

FIG 2 (a) Absorption spectra of SWNTs functionalized with PM (bacteriorhodopsin) kept in broadband light for varying time intervals from 1 to 4 hours (top to bottom). Inset shows absorption spectra of bacteriorhodopsin with its maximum peak at 570 nm. (b) showing optically induced rate of aggregation in PM functionalized SWNTs as compared to pristine SWNTs and PM functionalized SWNTs kept in the dark. Enhanced aggregation is seen for green light as compared to red illumination.

Fig. 2 illustrates absorption spectra of supernatant of optically aggregated PM functionalized SWNT dispersions. Pristine dispersions of SWNT do not show any change in concentration over the course of these experiments and very slow rate of aggregation as plotted is observed for PM functionalized SWNTs kept in dark; however, non linear, rapid aggregation is observed for the PM functionalized SWNT under a broadband illumination. Absorption spectra of SWNTs in aqueous solution show characteristic Van Hove singularities with peaks in VIS/NIR region due to semi-conductor SWNTs, and background absorption in UV and visible frequency is essentially because of plasmonic resonance of metallic SWNTs. Absorption spectra show a consistent decrease in concentration of SWNTs with illumination for different time intervals from one to four hours. Since, NIR absorption band of SWNTs are directly related to the band gap energy of the semiconducting nanotubes.[25] Changes in NIR band of absorption spectra of the separated supernatant indicate selective binding of the PM with selective diameters of SWNTs.

To understand the dependence on frequency of illumination on peak absorption of the PM, phenomenon was further studied using red and green band pass filters. Significantly enhanced aggregation is observed for green filter compared to red for normalized optical intensity. Rate of aggregation under red and green illumination is found to be linear, as compared to logarithmic decrease in concentration, in case of broadband illumination. Absorption spectra of the pure PM (inset Fig. 2) with molar absorption coefficient $63,000^{-1}$ $cm^{-1}$ show broad band with $\lambda_{max}$ at 570 nm. In the range 350-470 nm, PM absorption does not exceed 20% of $\lambda_{max}$, whereas absorption significantly decreases beyond 600 nm into visible NIR frequency. Significantly higher rate of aggregation for green illumination as compared to red emphasizes the dependence of phenomenon on the absorption of PM.

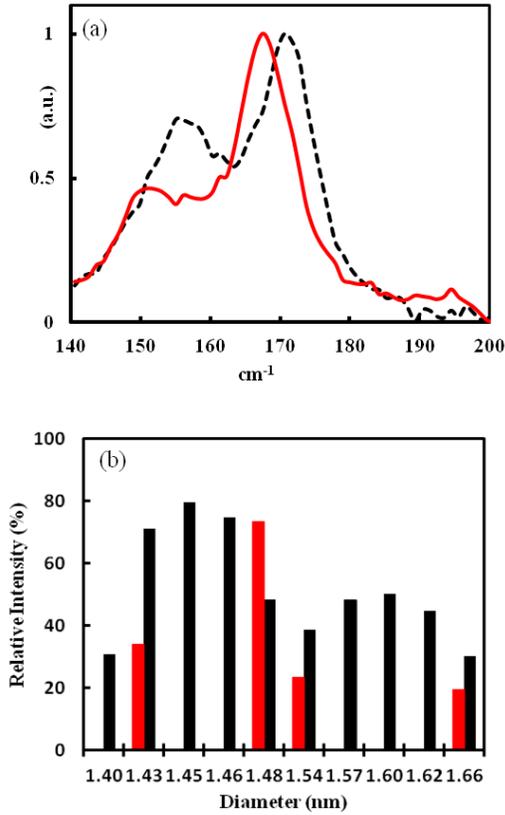

FIG 3 (a) RBM showing preferential binding of PM with SWNTs of specific diameters (red line) as compared to pristine SWNTs (black line). (b) Histogram as plotted using multiple peak fit in RBM showing relative enrichment in SWNTs of these specific diameters.

Samples were then centrifuged for 10 minutes at 5,000 rpm (Eppendorf® Mini Spin Centrifuge) to carefully separate the supernatant and aggregated floc from the solution. Raman spectrometer, labRAM, with 632 nm laser is used to further investigate the specific binding of the PM functionalized SWNT complexes. Fig. 3 (a) shows Radial breathing modes (RBM) of pristine SWNTs and that of aggregated floc, separated after 240 minutes of broadband visible illumination. RBM are due to a radial motion of atoms perpendicular to the axis and the resonant frequency is inversely related to the diameter of the tube by $\omega_{RBM} = (\alpha_{RBM}/d) + \alpha_{bundle}$. Where, d is the diameters of nanotubes and $\alpha_{RBM}$ (223.5 cm$^{-1}$), $\alpha_{bundle}$ (12.5 cm$^{-1}$) are constants, respectively.[25] Histogram, as shown in Fig. 3 (b) has been plotted using RBM data by fitting multiple Lorentzian peaks, representing individual diameters of control pristine SWNT solution and aggregated floc. The plot shows enrichment of SWNTs in aggregated floc specifically with as calculated diameter of 1.45 nm. *G* and *G'* band as shown in Fig. 4, show specific binding and strong interaction between the bR and SWNTs to form stable functional complexes. *G* band in pristine SWNT shows narrow $G^+$ Lorentzian peak at 1590 cm$^{-1}$ and broad $G^-$ peak around 1540 cm$^{-1}$ due to TO (circumferential) and LO (axial) modes.[26] Fig. 4 (a) shows increase in $G^-$ peak of functionalized SWNT as compared to pristine SWNTs. An increase in $G^-$ indicates an increase in metallic properties in separated, PM functionalized SWNTs. As reported, functionalized SWNT may also show changes in $G^-$ band due to interaction between electron donors/acceptor molecules.[27] Hence, an increase in $G^-$ band of a separated PM functionalized SWNT may be either due to specific binding as shown in the RBM mode or due to charge transfer and interaction between optically active bacteriorhodopsin and SWNTs. Shifts in G' band has been reported due to surface binding and functionalization of SWNTs. Hence, the red shift in G' band as shown in Fig. 4 (b) indicates strong binding and surface interaction between the PM and SWNTs. Moreover, as confirmed by no discernable changes in the D band (Fig. 4 (b) inset), this non-covalent functionalization does not affect the Sp$^2$ carbon hybridization; thus, retaining pristine electro-optical and charge transport properties of SWNTs.

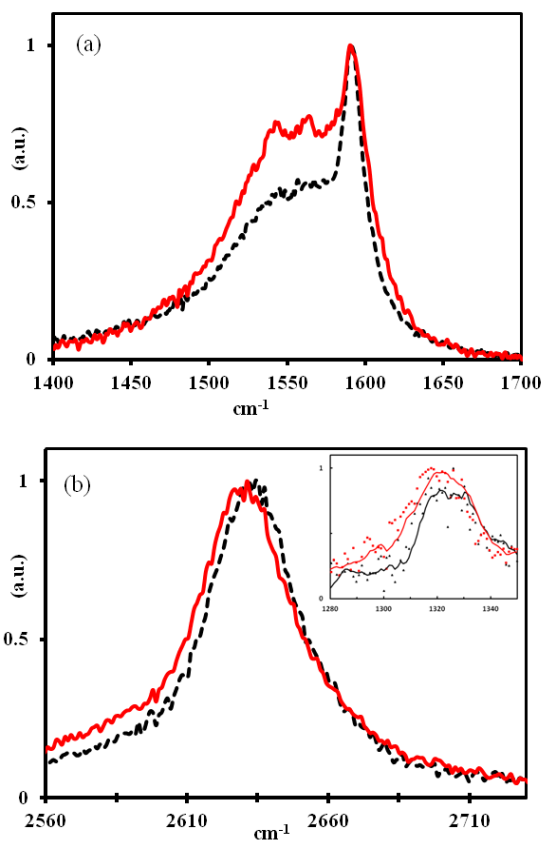

FIG 4 (a) *G* band Raman spectra showing increase in $G^-$ band ~ 1540-1560 cm$^{-1}$ in optically aggregated SWNTs functionalized with PM (red solid line) as compared to control pristine SWNTs (black dotted line). (b) *G'* band shows slight red shift indicating strong binding in PM functionalized SWNTs. However, (Inset) shows no discernible changes in the defect *D* band of SWNTs in the bio-nano hybrid complex.

Understanding fundamental interactions in solutions and using controlled external forces is significant for bottom up directed assembly of bio-nano hybrid devices. Surely, more research is needed to better understand complex interactions in SWNT functionalized with an optically active PM. However, our results confirm optically induced aggregation and consequent separation of the PM functionalized SWNT of selective diameters from solution. Enhanced aggregation is observed for optical frequency close to peak absorption frequency of bR, indicating strong interaction between the optically active PM and SWNT. The aggregated floc shows enrichment of SWNTs of specific diameters, indicating selective binding of PM with specific SWNTs. The stable bio-nano hybrid complexes based on SWNTs functionalized with optically active PM indicate promising potential electro-optical applications. Moreover, the PM functionalized SWNT provides an opportunity to better understand interactions in colloidal solutions in optically active, donor-acceptor systems. The functionalized complex also provides stable system for fundamental studies on charge transfer and interaction between bio-nano hybrid interfaces in solution. Optically induced aggregation of a selective PM functionalized SWNTs with specific strong binding certainly promises potential application in optically induced separation of other functionalized nanoparticles from solution, too. In conclusion, we report optical induced aggregation and separation of stable bio nano hybrid complex of the PM, functionalized preferentially with selective diameters of SWNTs.

Acknowledgement: Authors are indebted to Professor Oesterhalt, Max planck, Germany for the gift of Halobacterium S9 strain. We also thank G. Madhusudhan for scientific inputs and Ramanujan fellowship (*SR/S2/RJN-28/2009*) and funding agencies DST (*DST/TSG/PT/2012/66*), Nanomission (*SR/NM/NS-15/2012*) for generous grants.